\documentclass[aapm,mph,reprint,epsf,graphicx,showpacs]{revtex4-1}
\usepackage{graphicx}

\usepackage{amssymb}
\usepackage{amsmath}

\begin{document}


\title{
A molecular dynamics simulation of DNA damage induction by ionizing radiation
}

\author{
Ramin M. Abolfath, David J. Carlson, Zhe J. Chen, Ravinder Nath
}

\affiliation{
Department of Therapeutic Radiology, Yale University School of Medicine, New Haven, CT, 06520-8040
}
\noindent

\date{\today}

\begin{abstract}

\noindent \textbf{Purpose:} We present a multi-scale simulation of early stage of DNA damages by the indirect action of hydroxyl ($^\bullet$OH) free radicals generated by electrons and protons.

\noindent \textbf{Methods:} The computational method comprises of interfacing the Geant4-DNA Monte Carlo with the ReaxFF molecular dynamics software. A clustering method was employed to map the coordinates of $^\bullet$OH-radicals extracted from the ionization track-structures onto nano-meter simulation voxels filled with DNA and water molecules.

\noindent \textbf{Results:} The molecular dynamics simulation provides the time evolution and chemical reactions in individual simulation voxels as well as the energy-landscape accounted for the DNA-$^\bullet$OH chemical reaction that is essential for the first principle enumeration of hydrogen abstractions, chemical bond breaks, and DNA-lesions induced by collection of ions in clusters less than the critical dimension which is approximately 2-3 \AA. We show that the formation of broken bonds leads to DNA base and backbone damages that collectively propagate to DNA single and double strand breaks. For illustration of the methodology, we focused on particles with initial energy of 1 MeV. Our studies reveal a qualitative difference in DNA damage induced by low energy electrons and protons. Electrons mainly generate small pockets of $^\bullet$OH-radicals, randomly dispersed in the cell volume. In contrast, protons generate larger clusters along a straight-line parallel to the direction of the particle. The ratio of the total DNA double strand breaks induced by a single proton and electron track is determined to be $\approx$ 4 in the linear scaling limit.

\noindent \textbf{Conclusions:} In summary, we have developed a multi-scale computational model based on first principles to study the interaction of ionizing radiation with DNA molecules. The main advantage of our hybrid Monte Carlo approach using Geant4-DNA and ReaxFF is the multi-scale simulation of the cascade of both physical and chemical events which result in the formation of biological damage. The tool developed in this work can be used in the future to investigate the relative biological effectiveness of light and heavy ions that are used in radiotherapy.
\end{abstract}

\maketitle

\section{Introduction}

Ionizing radiations (X/ $\gamma $-rays, $\alpha $-particles, and heavy ions) induces damages to DNA-molecules such as single-strand breaks (SSB), double-strand breaks (DSB) and base damages (BD) via complex processes of direct ionizations and/or indirect action by free-radicals with a ratio that is determined by the energy and the type of radiation source.

\noindent
In the indirect mechanism of radiation interactions, the radiation ionizes water molecules and creates neutral free-radicals and aqueous electrons (Ward 1988, Mozumder 1985, Oliveira 2012, Moiseenko 1998). The process involves the generation and diffusion of $^\bullet$OH radicals in biological and/or aqueous environments followed by chemical reactions that allow removal of hydrogen atoms from the DNA. This process is energetically favorable for $^\bullet$OH radicals as it forms a water molecule and fills the electronic shell by neutralizing its magnetic moment.

\noindent
Great efforts have devoted to the statistical modeling of the damage sites based on Monte-Carlo (MC) sampling of the radiation track-structure and clustering of ionizations, using empirical reaction rates and radiation scattering cross-sections (Aydogan 2008, El Naqa 2012, Moiseenko 1998, Friedland 1999, Terrisol 1990, Wilson 1994, Nikjoo 1994, 1995, 1997, Zaider 1984, Semenenko 2005, Bernal 2009, Bernal 2011, Francis 2011, McNamara 2012, Kalantzis 2012, Michalik 1993, Brenner 1992, Lea 1946, Stewart 2011). Along this line we propose a first principle quantum mechanical simulation using the molecular dynamics (MD) computational models to provide more information on the biological and chemical pathways of radiation-induced DNA damage. In the chemistry literature, MD have shown great success for applications such as enzymatic reactions (Antoniou 2004, Karplus 1990, Karplus 2002, Schwartz 2009) and drug design (Durrant 2011).

\noindent
In nano-dosimetry studies, see e.g., (Moiseenko 1998, Bernal 2009, Bernal 2011), molecular structure of DNA and chromosomes has been used for scoring direct and indirect hits. In these models geometrical volumes are constructed to represent assemblies of atoms, e.g., the DNA-base.
Alternatively atoms can be described by an effective and adjustable volume such that if energy deposition occurs in a volume occupied by assemblies of DNA-atoms (e.g., DNA bases) or water molecules in the hydration shell, a direct hit is recorded, and this particular deposition is removed from further consideration.

\noindent
For DNA damage caused by $^\bullet$OH-radicals, an energy deposition has to occur within an adjustable diffusion length from the DNA. Coarse-grained reaction-diffusion rate models (Nikjoo 1994, 1995, 1997, Zaider 1984) have been developed to describe the process. In the atomistic level, the calculation of the diffusion length requires combination of quantum mechanics (QM) and molecular mechanics (MM). Preliminary results using QMMM (Abolfath 2012) show $^\bullet$OH-radicals close to 1 nm from the surface of DNA can reach the DNA within a ps time scale and perform subsequent chemical reaction.

\noindent
The average diffusion distance of a hydroxyl radical in a cellular milieu has been measured to be about 6 nm (Roots 1975). This is approximately three times the diameter of the DNA double helix, which implies that the chance an $^\bullet$OH to damage the DNA decreases rapidly with distances beyond about 6 nm.  To draw an accurate conclusion on the diffusion length predicted by MD and for comparison with the experiments, a systematic calculation must be carried out along the previous work to complete the results.

\noindent
For the current problem of interest, we introduce an alternative computational approach that permits going beyond phenomenological models and volumetric approaches such as those presented in Refs. (Aydogan 2008, El Naqa 2012, Moiseenko 1998, Friedland 1999, Terrisol 1990, Wilson 1994, Nikjoo 1994, 1995, 1997, Zaider 1984, Semenenko 2005, Bernal 2009, Bernal 2011, Francis 2011, McNamara 2012, Kalantzis 2012, Michalik 1993, Brenner 1992,L eaBook) as well as QMMM approaches such as Ref. (Abolfath 2012).

\begin{figure}
\begin{center}
\includegraphics[width=1.0\linewidth]{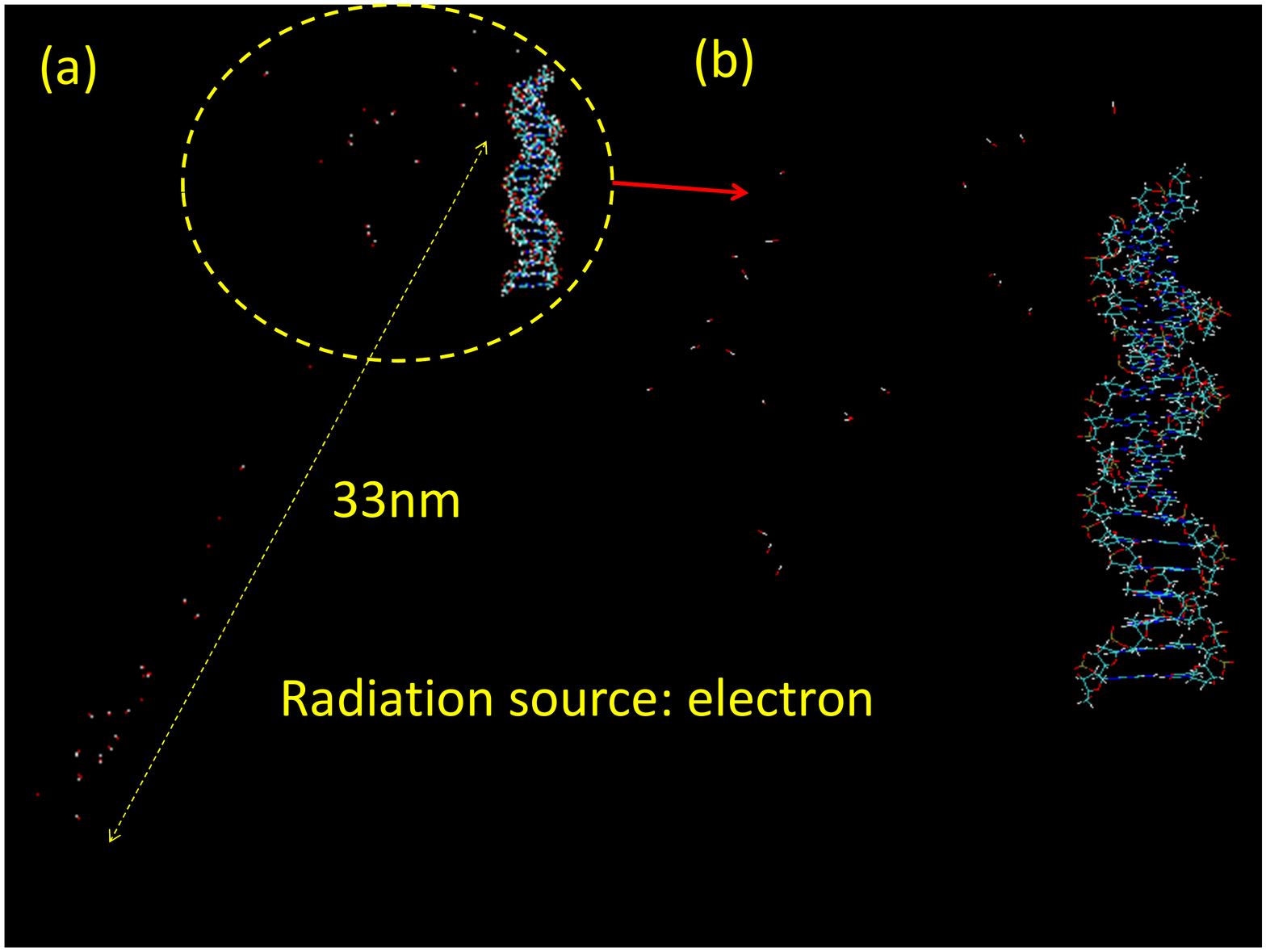}\vspace{-1.0cm}
\includegraphics[width=1.0\linewidth]{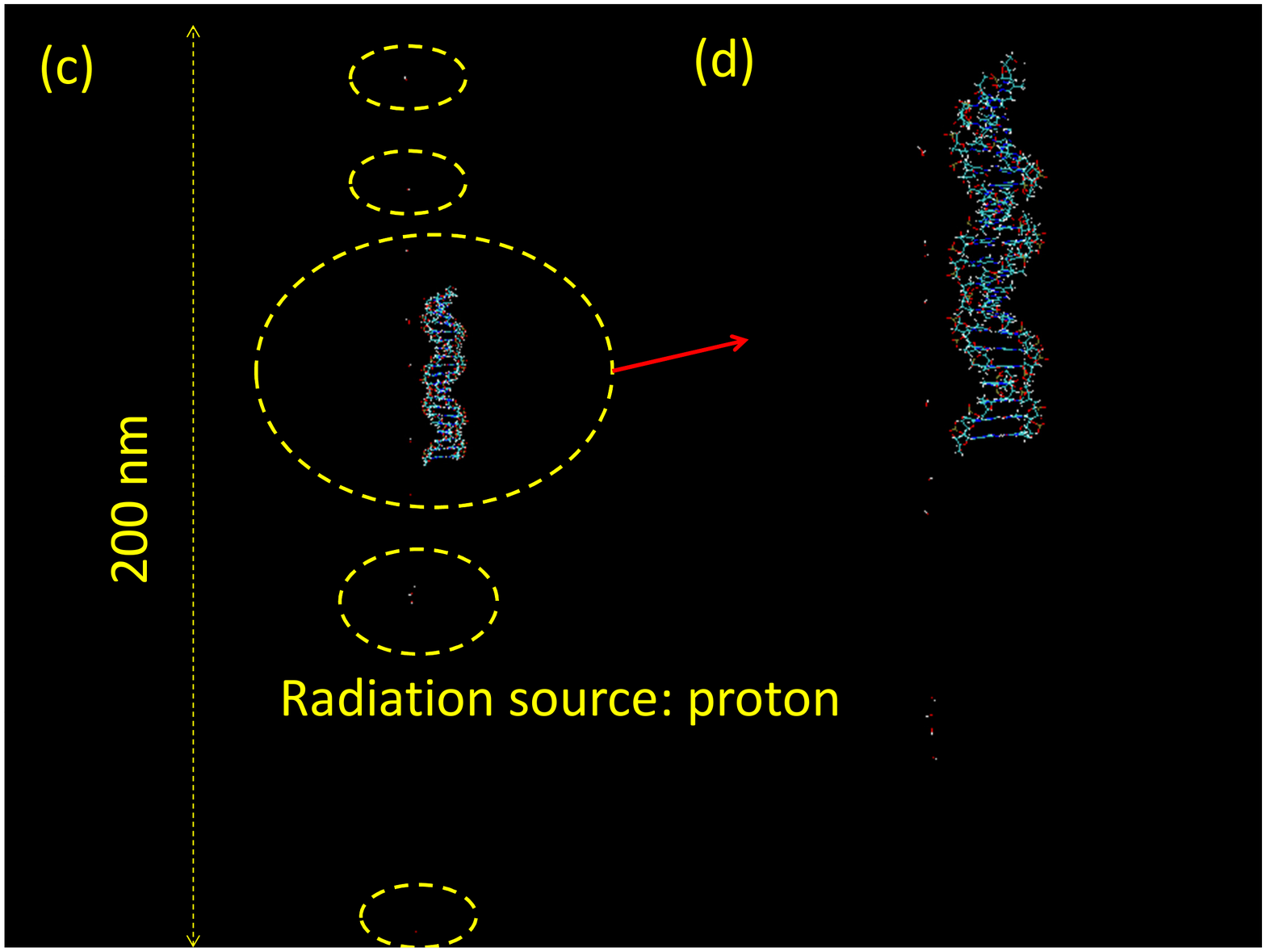}
\noindent
\caption{
(color online). A sample of ionization calculated by Geant4-DNA for a single-track of electron (a-b) and proton (c-d) in nano-meter scale. For illustration a segment of DNA parallel to the direction of central axis of the beam is added. Ions generated by the beam of electron are scattered in space. In comparison, ions generated by the beam of proton are in a straight line along the initial direction of proton. At the point of ionizations we introduce a diatomic $^\bullet$OH-radical with a random diatomic orientation.
}
\label{Fig0}
\end{center}\vspace{-0.5cm}
\end{figure}

\begin{figure}
\begin{center}
\includegraphics[width=1.0\linewidth]{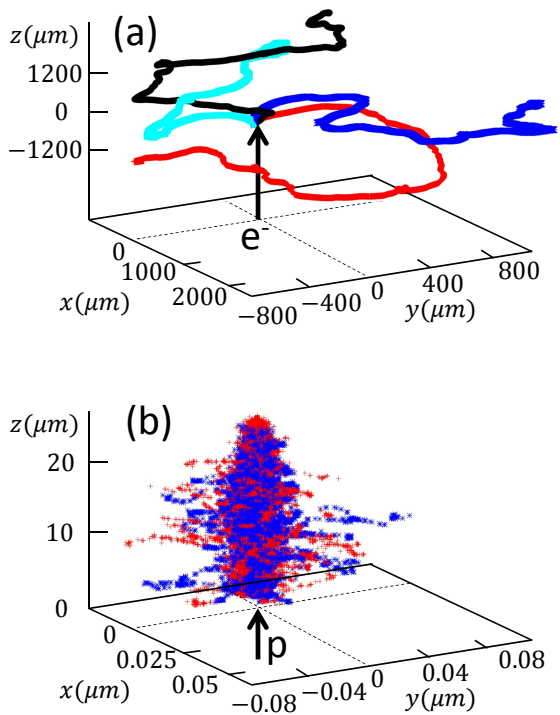}\vspace{-1.0cm}
\noindent
\caption{
(color online). Four ionization-tracks of 1 MeV electrons (a) and two ionization-tracks of 1 MeV protons (b) obtained from the Geant4-DNA Monte Carlo simulation in water with initial points at $x=y=z=0$. Colors represent tracks by a series of uncorrelated initial seeds in the Geant4 random number generator. The length is in micro-meter (mm) and the arrows show the initial direction of the beam.
}
\label{Fig1}
\end{center}\vspace{-0.5cm}
\end{figure}

\begin{figure}
\begin{center}
\includegraphics[width=1.0\linewidth]{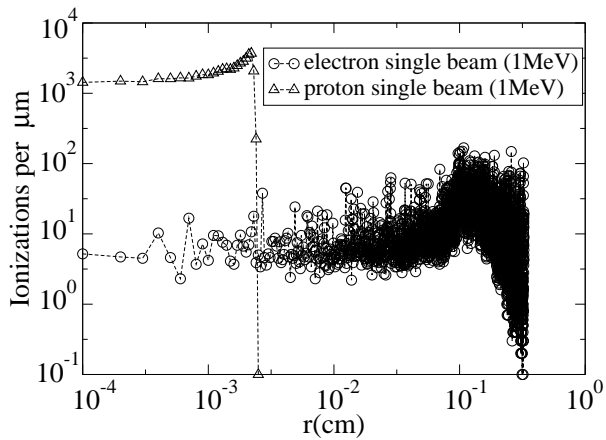} 
\noindent
\caption{
(color online). Track averaged radial density of ionizations as a function of distance from the source. Track averaging was performed over ten 1 MeV electrons and protons. The ionization density generated by protons shows a peak at the proton range, resembling the Bragg peak. The total number of ionizations per track for both electrons and protons is $\approx$50,000.
}
\label{Fig2}
\end{center}\vspace{-0.5cm}
\end{figure}

\begin{figure}
\begin{center}
\includegraphics[width=1.0\linewidth]{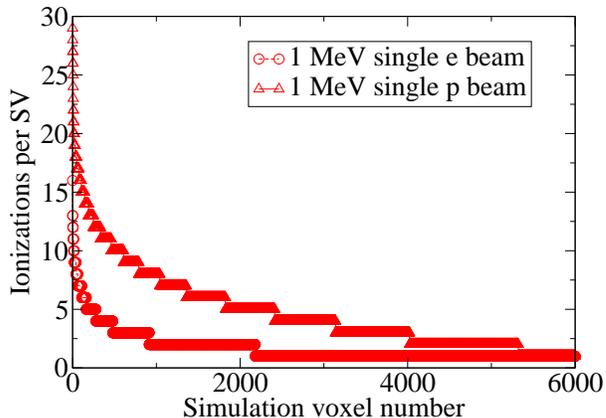} 
\noindent
\caption{
(color online). Histogram of ionizations distributed among nm-size SVs for a single electron and proton track (red tracks shown in Figs. \ref{Fig1} (a) and (b), respectively). The total number of SVs occupied by at least one ionization is 46,000 and 28,000 for an electron and proton, respectively.
}
\label{Fig3}
\end{center}\vspace{-0.5cm}
\end{figure}

\begin{figure}
\begin{center}
\includegraphics[width=1.0\linewidth]{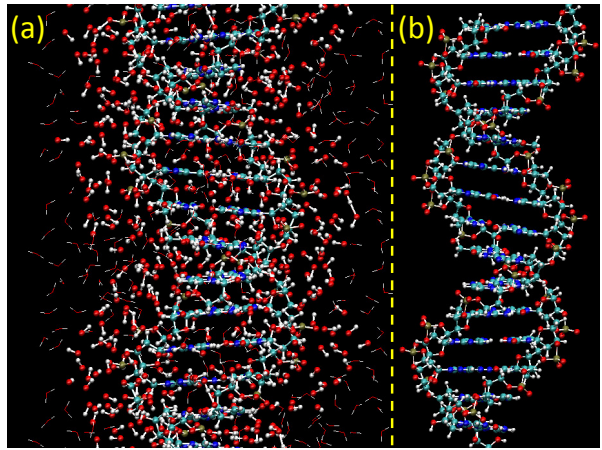} 
\noindent
\caption{
(color online). Final structure of DNA in solution after 10 ps MD simulation at room temperature. For clarity in visualization, in (b) the water molecules are removed from the identical computational box shown in (a).
Atoms color code: carbon, oxygen, nitrogen, phosphorus and hydrogen atoms are shown as cyan, red, blue, gold and white, respectively.
}
\label{Fig4}
\end{center}\vspace{-0.5cm}
\end{figure}

\begin{figure}
\begin{center}
\includegraphics[width=1.0\linewidth]{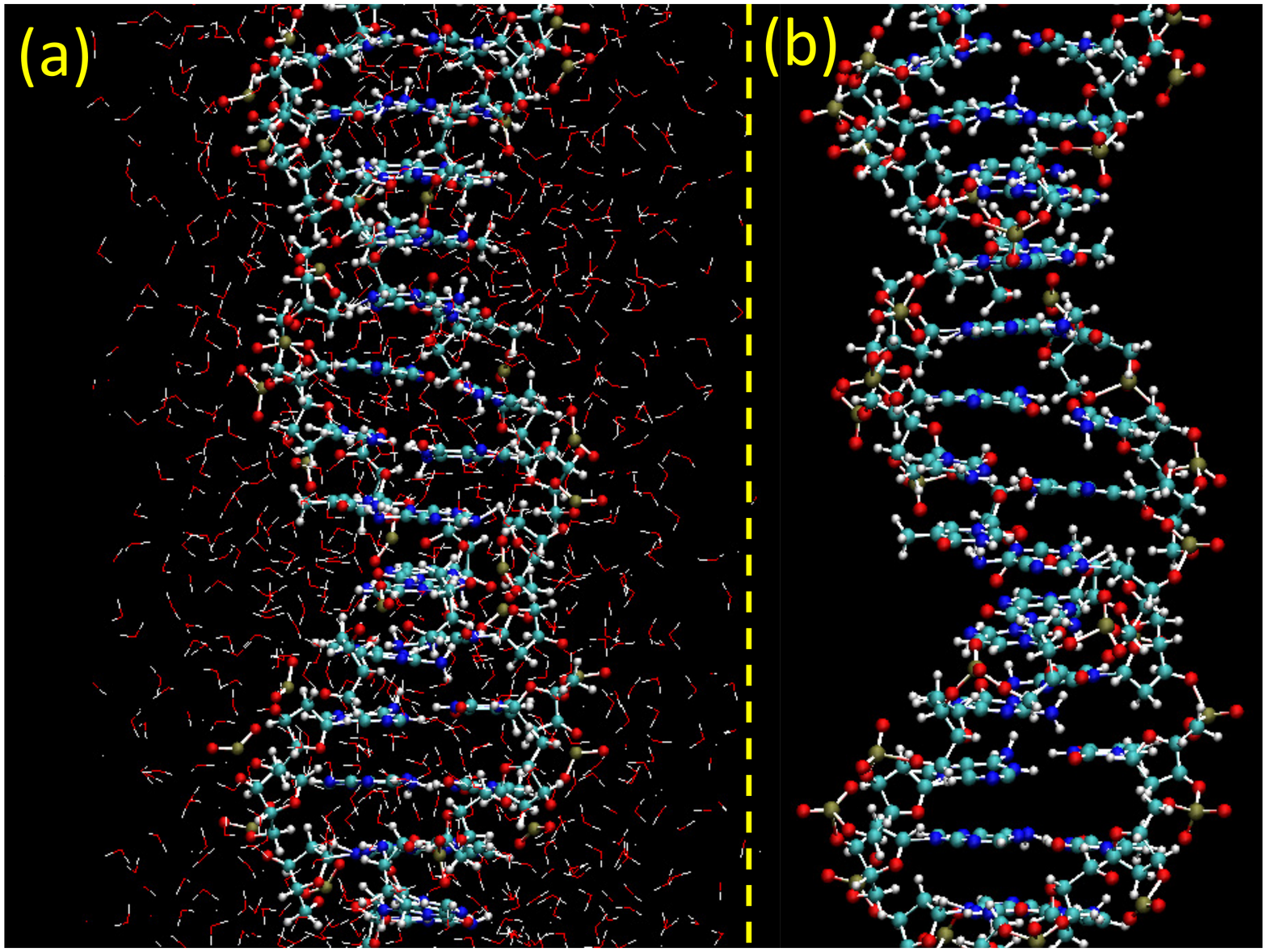} 
\includegraphics[width=1.0\linewidth]{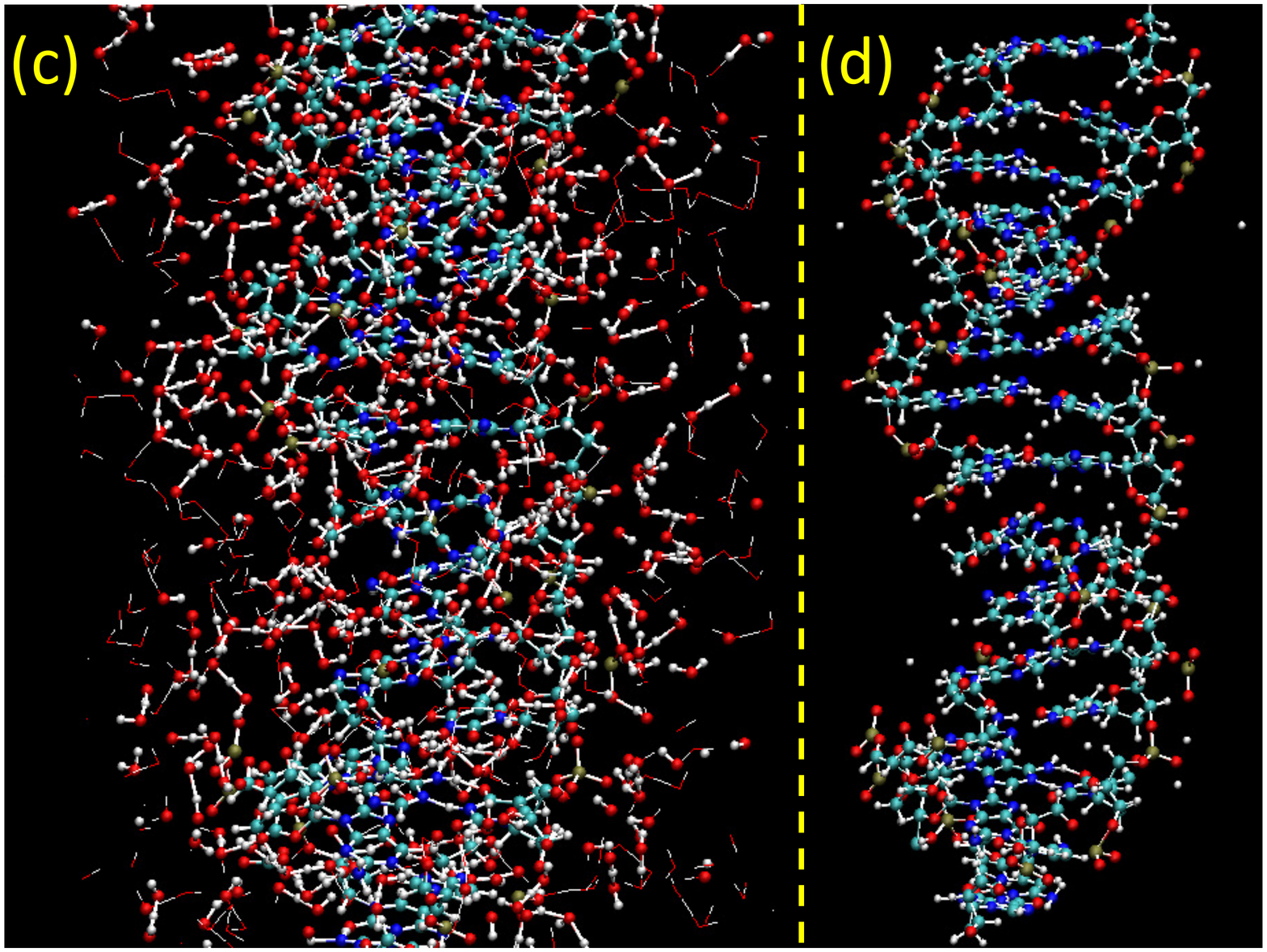} 
\noindent
\caption{
(color online). (a) Initial distribution of water molecules and randomly generated small pockets of $^\bullet$OH-radicals surrounding the DNA (\textit{t}=0). The dimension of the MD computational box (SV) is 2.6 $\times$ 2.6 $\times$ 6 nm${}^{3}$. (b) For visual clarity, water molecules and $^\bullet$OH-radicals are removed from the identical computational box shown in (a). (c) Final configuration of DNA at \textit{t=10} ps in water and in the presence of $^\bullet$OH-radicals. (d) Same as (c) after deleting water and $^\bullet$OH molecules. The white dots in the background are abstracted hydreogens from DNA. The color code used to label the atoms is cyan, red, blue, gold and white for carbon, oxygen, nitrogen, phosphorus and hydrogen, respectively.
}
\label{Fig5}
\end{center}\vspace{-0.5cm}
\end{figure}

\begin{figure}
\begin{center}
\includegraphics[width=1.0\linewidth]{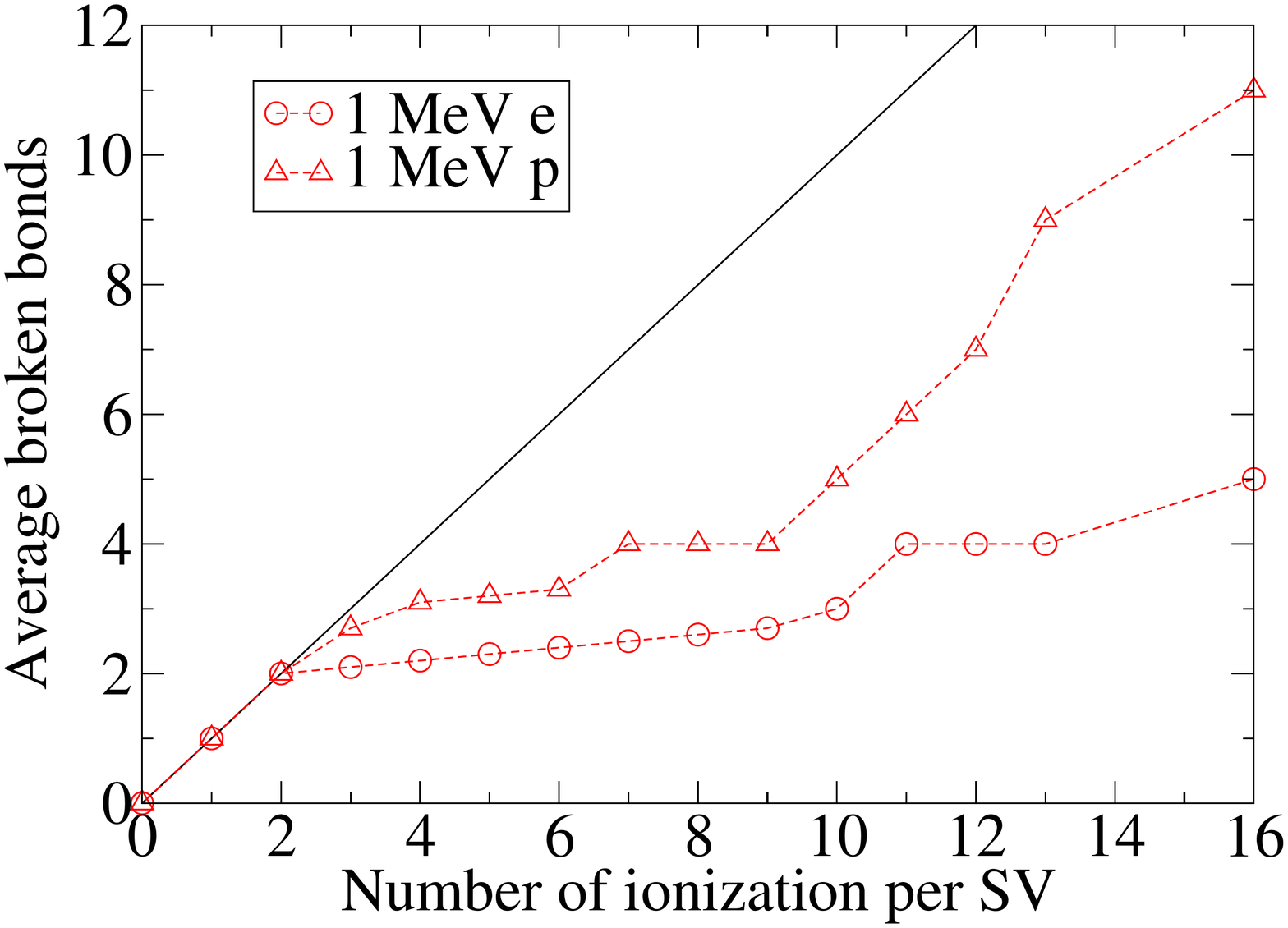} 
\noindent
\caption{
(color online). Average number of DNA broken bonds, $\bar{{\rm L}}_{{\rm N}}^{\alpha } $, as a function of number of ionizations \textit{N} per SV, calculated for a single electron (e) and proton (p) track (red tracks shown in Fig. \ref{Fig1} (a) and (b), respectively). Solid line shows the linear scaling limit, $\bar{{\rm L}}_{{\rm N}}^{\alpha } =N$ where one ion generates one DNA broken bond, neglecting sub-critical clusters. The deviation from linear scaling for electron is clearly more pronounced than for proton.
}
\label{Fig6}
\end{center}\vspace{-0.5cm}
\end{figure}

\section{Materials and Methods:}
A selective sampling of the DNA damage caused by the ionization track structure is performed by partitioning the cm-size MC volume into a three dimensional lattice of 10$^{19}$  simulation voxels (SVs) with dimension 2.6 $\times$ 2.6 $\times$ 6 nm$^3$. In each SV, we placed a fragment of DNA with the helical axis parallel to the beam direction and fill the empty space of the SV with water molecules.  Geant4-DNA MC simulation is used to calculate the spatial distributions of ions in and among SVs and to characterize the dependence of DNA-damage pattern on the beam source and quality.

\noindent
To generate the coordinates of the ions on a nano-scale level, we employ the GEANT4-DNA extension of the GEANT4 Monte Carlo simulation toolkit, version 9.6.p01-64bits, (Incerti 2010a, 2010b). The GEANT4-DNA package allows the event-by-event simulation of the particle shower produced during the transport of electrons and protons in a continuous model of liquid water.

\noindent
In the current version of Geant4-DNA, the static structure of the water molecules is taken from the scattering cross sections. However, the molecular representation of water included in the physics lacks the dynamical aspects needed for studying the time-evolution of ionization.
Moreover the DNA structure is introduced as a target volume inside a water phantom, pretending that the target volume is filled by DNA (Bernal 2011). In our approach, we assume that at the location of ionized water, an $^\bullet$OH free radical is created. By engaging MD simulation, we calculate the dynamical trajectory and the energy-landscape of the DNA molecule in the presence of water and $^\bullet$OH-radicals on-the-fly and hence the first-principle enumeration of DNA lesions driven by cascade of chemical reactions can be carried out. Here, and throughout this work, a DNA lesion refers to a single SSB or BD. The energy-landscape of the system is the total potential energy of DNA and $^\bullet$OH, calculated as a function of reaction distance, e.g., the distance between DNA and the transferred hydrogen to $^\bullet$OH. This hydrogen transfers to $^\bullet$OH and form a water molecule. As a result of collective migration of hydrogen from DNA, cascade of chemical reactions such as  carbonyl- and hydroxyl-hole formation in the sugar-moiety rings, nucleotide-nucleotide hydrogen bond disruption, nucleotide structural damage and nucleotide-sugar-moiety bond breaking occur.

\noindent
We further investigate the dependence of DNA-$^\bullet$OH reactivity on the formation of hydrogen-peroxide compounds and the network of $^\bullet$OH$\cdots$$^\bullet$OH hydrogen bonded molecules.
We demonstrate the stability of these chemical complexes and show that they prevent free radicals from reaching the DNA molecule.
Such complexes tend to form for $^\bullet$OH-radicals in close proximity to each other, i.e., in sub-critical clusters with dimension less than a critical value with radius smaller than 2-3 \AA. For a pair of $^\bullet$OH radicals the radius of a sub-critical cluster is determined by the hydrogen-bond effective length as two radicals can form $^\bullet$OH$\cdots$$^\bullet$OH complex (here $\cdots$  represents hydrogen bond) or a covalent bond length if they form hydrogen peroxide, HO-OH. In ReaxFF-MD, formation of $^\bullet$OH$\cdots$$^\bullet$OH and HO-OH bonds are seen for a distance between two $^\bullet$OH radicals less than 2-3 \AA.

\noindent
Our approach for reactive MD consists of ReaxFF (van Duin 2001, Chenoweth 2008) implemented in Large-scale Atomic/Molecular Massively Parallel Simulator, LAMMPS, version Jan. 2011 (Plimpton 1995, 1997).
In regard to the software performance in terms of computing time and memory usage, a modest computer with 4GB RAM is sufficient to run ReaxFF-MD program embedded in LAMMPS. The program is fast and efficiently parallelized. The speed depends on the time steps and varies between few hours to few days.
The consistency and accuracy of ReaxFF with local density functional calculation is checked systematically by employing ab-initio Car-Parrinello MD (Car 1985, Hutter 2008).
In ReaxFF the atomic interactions are described by the reactive force field potential (van Duin 2001). ReaxFF is a general bond-order dependent potential that provides accurate descriptions of bond breaking and bond formation. Recent simulations on a number of hydrocarbon-oxygen systems (Bagri 2010), and organic molecules (Abolfath 2011, Monti 2012) showed that ReaxFF reliably provides energies, transition states, reaction pathways and reactivity trends in agreement with ab-initio calculations and experiments.

\noindent
In our method, the spatial distribution of the ionized water molecules and $^\bullet$OH-radicals are obtained by post-processing of the track structures calculated by Geant4-DNA. A series of statistically uncorrelated single particle tracks are generated and used for the sampling of the DNA lesions by changing the random number generator seeds. We specifically focus on single electron/proton beams with initial energy of 1 MeV as an illustration of the methodology. We obtain the coordinates and spatial distribution of water ionizations in a cm-size phantom for each track. For the statistical sampling of $^\bullet$OH in SVs, for each track we enumerate the number of ions per SV, \textit{N}, and the number of SVs, K${}_{N}$, for a given \textit{N}.

\noindent
The chemical interaction between hydroxyl radicals and DNA and the scoring of hydrogen abstractions, chemical bond breaking, and DNA lesion formation is simulated by ReaxFF-MD. We fill the SVs with DNA and water molecules corresponding to the density equal to 1 g/cm${}^{3}$ using the computational tools available in GROMACS (Berendsen 1996, Lindahl 2001, Van Der Spoel 2005, Hess 2008, Abolfath 2012).
GROMACS is a versatile package to perform classical molecular dynamics. It simulates the Newtonian equations of motion for systems with hundreds to millions of particles.

\noindent
An interface between the output of Geant4-DNA with the input of ReaxFF-MD (Abolfath 2011, van Duin 2001) was constructed to convert the coordinates of ionization events obtained in Geant4-DNA to generate the coordinates of diatomic structure of $^\bullet$OH-radicals used in ReaxFF-MD at the ionization points with random angular orientations with respect to the DNA-molecule, as shown in Fig.~\ref{Fig0}.

\noindent
A series of ReaxFF-MD were performed to calculate the average number of DNA lesions per SV per beam for a given \textit{N} ionizations. A DSB consists of two or more SSB on opposite strands of the DNA that are spatially and temporally close (Ward 1988). As shown in Fig.~\ref{Fig0}, we construct a large fragment of a solvated 950-atom classical Watson-Crick DNA-strand (Munteanu 1998), consisting of 15 base-pairs, with approximately 1,100 water molecules added using GROMACS (Berendsen 1996, Lindahl 2001, Van Der Spoel 2005, Hess 2008, Abolfath 2012) along the principal axis of the SV, the \textit{z}-axis.

\noindent
The vertical dimension of the SV and the length of the DNA molecule are optimized to balance between the minimum size of a DNA to fit one DSB per DNA-length and the largest fragment of DNA including the surrounding water molecules, allowable in the MD simulation. The diffusibility of $^\bullet$OH radicals must be taken into account to estimate a reasonable cut-off on the SV lateral dimensions. When the lateral size of SV increases, the number of water molecules surrounding the DNA increases and induces an unnecessary increase of the simulation time. Consequently, we
chose the lateral dimension of the SV equal to 2.6 nm.

\noindent
To minimize the computational time, it is favorable to choose SVs as small as possible. However, by choosing very small SVs, we may lose layers of water molecules that are within the $^\bullet$OH diffusion length. Therefore, an optimal value of the lateral dimension of the SV is determined by optimizing the number of water molecules surrounding the DNA to incorporate properly the diffusion of $^\bullet$OH-radicals within ps time-scale.
We use the constructed molecular structure as a replica for all SVs in the Geant4-DNA volume. $^\bullet$OH-radicals are added in the position of ionization coordinates obtained from Geant4-DNA with random orientation relative to the DNA. These simulations were performed using periodic boundary conditions in a canonical moles, pressure and temperature (NPT) ensemble, with a Nose-Hoover thermostat for temperature control and a time step of 0.25 fs and run MD for 50 ps, as described in Ref. (Abolfath 2011). Within this period of time, $^\bullet$OH radicals interact with DNA to induce DNA lesions, i.e., SSB and  BD, and subsequently DSB.

\noindent
Two different sets of MD simulation were performed: in the presence and absence of $^\bullet$OH-radicals. We follow the molecular dynamics of $^\bullet$OH radicals in the presence of DNA and water using molecular time-evolution built in the ReaxFF-MD. Therefore, all events in our approach are accounted for as indirect damage by the $^\bullet$OH free radical. The relevant time-scales for the sequence of simulations, consistent with the experimental data (El Naqa 2012,Moiseenko 1998, Sies 1993) ranges from atto- to nano-seconds, relevant to MC and MD events, respectively (Mozumder 1985). From the output of the MD calculation, we count the number of broken DNA bonds resulting in DNA lesions and DSBs for each track. The above procedure is then repeated to obtain the final results that are averaged over all simulated tracks.

\section{Results:}
Fig. \ref{Fig1} shows samples of the ionization-tracks of 1 MeV electrons (a) and protons (b) obtained from the Geant4-DNA Monte Carlo simulation in a continuous model of water. The maximum beam energy in the current version of Geant4-DNA is 1 MeV. The arrows show the initial direction and origin of the particles. Each point represents a single ionization event. Although the total number of ionizations produced by a single 1 MeV electron and proton track is approximately the same (about 50,000), the range of a 1 MeV proton is clearly shorter than the range of a 1 MeV electron due to the greater density of ionizations along the proton track. Because the electron undergoes stronger lateral scattering by the medium compared to the proton, we observe a strong divergence of the ionization tracks of electrons in space compared to protons. Moreover, a large fraction of the ions produced by the electrons are in the negative direction relative to the direction of the incident electron, indicating that the number of back-scattering events for electrons is much greater than for protons.

\noindent
Fig. \ref{Fig2} illustrates the radial density of ionizations averaged over ten tracks and tallied into one micro-meter radial slices. The distance from the source, r, is given for the ionization tracks shown in Fig. \ref{Fig1}. The average ionization range of 1 MeV electrons and protons is 3000 and 25 $\mu$m, respectively. The calculated ionization range of electron is consistent with the NIST-reported CSDA range of a 1 MeV electron of $\approx$ 4 mm. The latter includes all excitations including ionization events.
Proton ionization density shows a peak in the tail of the track, resembling the Bragg peak. The average depth of the range of the electrons occurs at a distance two orders of magnitude larger than the average proton range because the mass of the electron is three orders of magnitude smaller than the proton. The maximum number of ionizations per micro-meter length per track accounted for the proton beam is three orders of magnitude larger than that for the electron beam. The calculated total ionization energy deposited by the electrons and protons per track is 660 and 640 keV, respectively. Hence, we find the electron and proton stopping powers averaged over the entire track to be approximately 0.22 and 26.6 keV/$\mu$m, respectively, which is consistent with published values (Bernal 2011).

\noindent
Fig. \ref{Fig3} shows a typical distribution of ionizations among 2.6 $\times$ 2.6 $\times$ 6 nm${}^{3}$ MD-SVs calculated for a single 1MeV electron and proton track (red tracks shown in Fig. \ref{Fig1} (a) and (b), respectively). There are 10${}^{19}$ SVs that cover the three-dimensional structure of the cm-size MC volume. The number of ions per SV per track, \textit{N}, is extracted from post-processing of the output of Geant4-DNA. \textit{N} is an integer number that ranges from zero to \textit{N$_{\max}$}. Because a single particle track deposits ions in only a small fraction of the SVs, the distribution is highly sparse, i.e., there is a large number of SVs with \textit{N=0}. As we defined \textit{K${}_{N}$} is the number of the SVs that are occupied by \textit{N} ions. This quantity partitions the SVs based on the number of ionizations. As pointed out above, the total number of ionizations deposited by the beam of electron is approximately the same as the beam of proton.  For clarity, the number of ionizations shown in the \textit{x}-axis of Fig. \ref{Fig3} is truncated at 6,000. The areas under the curves are the total number of ionizations.

\noindent
Figs. \ref{Fig4} and \ref{Fig5} show the structure of the DNA and water molecules in the ReaxFF-MD computational box. In SVs with zero ionization (\textit{N=0}), we do not expect any DNA-damage. However, validation of the ReaxFF-MD results can be demonstrated by performing molecular dynamics for this situation. The simulation illustrated in Fig. \ref{Fig4} describes equilibration of the solvated DNA-strand at T=300K for 50 ps. We found that during this time-frame the DNA retained its overall helical configuration, indicating that ReaxFF retains the overall structural integrity of the DNA over such time-frames and that reactive events observed during exposure to $^\bullet$OH radicals, as shown in Figs. \ref{Fig5} (a-d), can indeed be associated with the radical reactivity.

\noindent
Fig. \ref{Fig5} (b) illustrates the initial DNA structure used in MD simulation. During the simulation time, chemical reactions between $^\bullet$OH-radicals and DNA lead to DNA hydrogen abstraction. This can be observed as white dots in the background of Fig. \ref{Fig5} (d). In the final state, i.e., Figs. \ref{Fig5} (c) and \ref{Fig5} (d), a pronounced separation between bases and backbone is visible. Strong distortion in base stacking with large holes are clearly seen. Large separation between bases and the backbone as well as the stretched backbone illustrating a collective base-damage, and large number of single-strand breaks, are visible. As a result of chemical reactions, the charge distribution on the DNA is dramatically altered and the helical-structure becomes unstable. Consequently, a series of bonds are broken. In Fig. \ref{Fig5} (d) we observe number of missing links between atoms, accounted for broken covalent bonds. Hence, comparing Figs. \ref{Fig5} (b) and (d), we can clearly count the number of missing covalent bonds accounted for DNA lesions, as each broken covalent bond corresponds to one DNA lesion. Moreover, we identify chemical pathways for carbonyl- and hydroxyl-hole formation in the sugar-moiety rings, nucleotide-nucleotide hydrogen bond disruption, and nucleotide structural damage throughout a sequence of intermediate events such as DNA-backbone and -base hydrogen abstraction by $^\bullet$OH-radicals followed by nucleotide-sugar-moiety bond breaking.

\noindent
Figure \ref{Fig6} shows the average number of DNA broken bonds, $\bar{{\rm L}}^{\alpha }_{{\rm N}} {\rm \; }$per SV, as a function of the number of ionizations \textit{N} per SV and the particle-type, $\alpha $=\{e, p\}, calculated by ReaxFF-MD for the red-tracks in Fig. \ref{Fig1}. In the limit of small \textit{N}'s, and for both electron and proton, $\bar{{\rm L}}_{{\rm N}}^{\alpha } {\rm \; }\approx {\rm N}$. However, a deviation from linear scaling emerges for \textit{N = 3} due to the formation of sub-critical clusters of ions. The results obtained directly from MD simulation show that the ions localized in these clusters can form only a single broken bond in DNA, hence they suppress biologically effectiveness of the beam as argued by Lea (Lea 1946).

\noindent
The number of SVs, \textit{K${}_{N}$}, for a given \textit{N} can be calculated from Fig. \ref{Fig3}. For all tracks of electrons considered in this study, we find that 95\% of the ions contribute to single ionization per SV, i.e., the number of SVs that are occupied only with a single ion, \textit{N=1}, forms 95\% of the population of the ions.
This ratio drops to 80\% for the proton track. Because the DNA broken bond induced by a single ion is either on one side of DNA-backbone or on DNA-base, 95\% and 80\% of the ions in the track of a single electron and proton, respectively, do not participate in DSB formation. Equivalently 5\% and 20\% of the electron and proton track, respectively, populate SVs with at least \textit{N = 2} and hence they can potentially participate in DSB formation.

\noindent
For \textit{N=2}, the number of SVs containing sub-critical clusters is small, hence $\bar{{\rm L}}_{{\rm N}={\rm 2}}^{\alpha } {\rm \; }\approx {\rm 2}$. Considering only contribution of the \textit{N=2}-ions yields ${\rm K}_{{\rm 2}}^{{\rm p}} {\rm /K}_{{\rm 2}}^{{\rm e}} {\rm \; }\approx {\rm \; 1.02}$ where ${\rm K}_{{\rm 2}}^{{\rm p}} {\rm \; (K}_{{\rm 2}}^{{\rm e}} {\rm )}$ is the population of the SVs with two ions generated from a single particle track of proton (electron).
The total number of DSBs generated by a beam of ionizing radiation for a single track structure was calculated by convoluting the average number of lesions induced by a given number of ions, \textit{N}, per SV, $\bar{{\rm L}}_{{\rm N}}^{\alpha } $, and their population in SVs, ${\rm K}_{{\rm N}}^{\alpha } $:

\begin{eqnarray}
{\rm DSB}_{\alpha } =\sum _{N=0}^{N_{\max }^{\alpha } }f_{{\rm N}}^{\alpha }  {\rm \; }\bar{{\rm L}}_{{\rm N}}^{\alpha } {\rm \; K}_{{\rm N}}^{\alpha } .\begin{array}{ccc} {\begin{array}{ccc} {} & {} & {} \end{array}} & {} & 
\end{array}
\label{eq1}
\end{eqnarray}

\noindent
Here $f_{{\rm N}}^{\alpha }$ is a fraction of broken bonds propagating to DSBs and $\bar{{\rm L}}_{{\rm N}}^{\alpha } =\frac{1}{K_{{\rm N}}^{\alpha } } \sum _{i=1}^{K_{{\rm N}}^{\alpha } }L_{{\rm i}}^{\alpha }$ where $L_{{\rm i}}^{\alpha }$ is the number of broken-bonds in the \textit{i}th-SV, hence ${\rm DSB}_{\alpha } =\sum _{N=2}^{N_{\max }^{\alpha } }f_{{\rm N}}^{\alpha } \sum _{i=1}^{{\rm K}_{{\rm N}}^{\alpha } }{\rm L}_{{\rm i}}^{\alpha }   {\rm \; }.$ Note that $f_{{\rm N}}^{\alpha } =1/2$ if each pair of broken bond leads to a single DSB.  The number of DNA lesions in a SV, ${\rm L}_{{\rm i}}^{\alpha } $, depends on both the number of $^\bullet$OH-radicals, \textit{N}, and their spatial distribution. ${\rm L}_{{\rm i}}^{\alpha }$ fluctuates among ${\rm K}_{{\rm N}}^{\alpha }$ SVs because of the variation in the relative position of DNA and $^\bullet$OH radicals that influence the diffusion length and $^\bullet$OH-radicals among each other with the possibility of the occurrence of the sub-critical clusters.

\noindent
Calculation of Eq.\eqref{eq1} requires a series of MD simulations. For example, for a single 1 MeV electron and proton track (red track shown in Figs. \ref{Fig1} (a) and (b), respectively), it requires \textit{K${}_{N}$ = 1, 2, 3, 7, 7, \dots } and \textit{K${}_{N}$ = 29, 58, 84, 129, 147, \dots } ReaxFF calculation, both corresponding to \textit{N = 16, 13, 12, 11, 10, }\dots  (see Fig. \ref{Fig3}). Clearly K${}_{N}$ increases as \textit{N} approaches to zero, hence performing ReaxFF-MD simulation for all K${}_{N}$ is a computationally expensive problem. To speed up the calculation, we perform ReaxFF-MD for SVs, selected  randomly from the beginning, middle, and the tail of the ionization tracks. We then calculate ${\rm L}_{{\rm N}}^{\alpha }$ and ${\rm K}_{{\rm N}}^{\alpha }$ by averaging over the SVs for given tracks following by a second averaging over the simulated tracks. The average DSB per track, hence can be calculated by

\begin{eqnarray}
\left\langle {\rm DSB}_{\alpha } \right\rangle =\sum _{N=0}^{N_{\max }^{\alpha } }{\rm \; }\left\langle f_{{\rm N}}^{\alpha } \bar{{\rm L}}_{{\rm N}}^{\alpha } {\rm \; K}_{{\rm N}}^{\alpha } \right\rangle  ,\begin{array}{ccc} {\begin{array}{ccc} {} & {} & {} \end{array}} & {} &
\end{array}
\label{eq2}
\end{eqnarray}
where $\left\langle {\rm ...}\right\rangle$ denotes track-averaging.

\noindent
The dependence of the total number of ionizations on the number of tracks is approximately linear, assuming that the ionization-tracks are statistically uncorrelated. Hence, the total DSBs, ${\rm DSB}_{_{\alpha } }^{{\rm tot}} \propto N_{tracks} $for small $N_{tracks} $. Here $N_{tracks}$ is the number of tracks. However, the dependence of the number of DSBs on the number of tracks for both electrons and protons deviates from linear scaling owing to the saturation of the ions and increase in the population of the sub-critical clusters. With increase of $N_{\rm tracks} $, the sub-linear behavior of ${\rm DSB}_{_{\alpha } }^{{\rm tot}} $emerges, hence it asymptotically saturates to a maximum value. The onset of DSB saturation for proton appears in smaller $N_{\rm tracks} $ compare to electron as can be anticipated from Fig. \ref{Fig1}. In the linear scaling limit, where a system of multi-tracks reduces to a single-track problem, we find that the ratio of total DSB yields induced by a single electron and proton track with initial energy of 1 MeV, DSB${}_{p}$/DSB${}_{e}$, is  4.
Note that in this limit $\left\langle {\rm DSB}_{\alpha} \right\rangle \approx {\rm DSB}_{\alpha}$ where in the left hand side $\left\langle {\rm DSB}_{\alpha}\right\rangle$ denotes DSBs averaged over multi-tracks and ${\rm DSB}_{\alpha}$ in the right had side denotes the average of DSBs calculated from a single track.

\noindent
The upper limit of DSB${}_{p}$/DSB${}_{e}$ can be calculated by neglecting the contribution of the sub-critical clusters. In this case we assume that a pair of ions generate one DSB, $f_{{\rm N}}^{\alpha } =1/2$, and the number of DNA lesions scale linearly with number of ionization, $\bar{{\rm L}}_{{\rm N}}^{\alpha } =N$. In Fig. \ref{Fig6}, the solid line shows such limit. It follows

\begin{eqnarray}
{\rm DSB}_{\alpha } =\sum _{N=0}^{N_{\rm max}^\alpha} {\rm mod}(\frac{N}{2}) {\rm \; K}_{{\rm N}}^{\alpha},
\label{eq3}
\end{eqnarray}
where mod\textit{(N/2)} is the integer part of \textit{N/2}. For \textit{N=1} the number of DSBs is identical to zero. For \textit{N=2}, Eq. \eqref{eq3} predicts ${\rm DSB}_{\alpha } ={\rm K}_2^{\alpha } $ since mod\textit{(N/2)=1}. Clearly because Eq.\eqref{eq3} neglect the sub-linear behavior of DSB due to the emerging of the sub-critical clusters, it over-estimates the contribution of the ion-pairs thus it leads to a higher values for DSB${}_{p}$/DSB${}_{e}$ than predicted by MD simulation of Eq.\eqref{eq1}. For the case of 1 MeV beams, Eq.\eqref{eq3} predicts DSB${}_{p}$/DSB${}_{e} = 4.4$ that is 10\% greater than the one calculated by Eq.\eqref{eq1}.

\section{Discussion and conclusion:}
The DNA double strand break is the most biologically relevant damage induced by ionizing radiation. To this end, we built an interface to export the output of Geant4-DNA to the ReaxFF-MD environment. The steps included (a) dividing a cm-size Geant4-DNA computational box into $\approx$10${}^{19}$ nm-size MD-SVs (b) construction of a molecular structure of SVs under irradiation by adding water molecules as well as diatomic structure of $^\bullet$OH-radicals at the ionization points with random angular orientations with respect to 15-bp Watson-Crick DNA molecules (c) calculation of inter- and intra-SV distribution and numbers of $^\bullet$OH-radicals and (d) performing MD to enumerate DNA lesions.

\noindent
The range of energy used in this study is relevant to the low energy portion of a proton beam that results in a Bragg peak and is used in radiotherapy to deliver highly conformal dose distributions to tumors. The maximum particle energy was 1 MeV for both protons and electrons and the ionization events were collected from both single-track and multi-track beams. This maximum energy corresponds to the maximum energy available in the current version of Geant4-DNA for electron. However, our approach can be extended in the future to more clinically-relevant electrons and protons with energy greater than 1 MeV.

\noindent
In MD simulation, we assumed that randomly distributed clusters of diatomic $^\bullet$OH radicals are the main source of hydrogen abstraction. We demonstrated formation of carbonyl- and hydroxyl-groups in the sugar-moiety create holes that grow up slowly between DNA-bases and DNA-backbone and the damage collectively propagates to DNA single and double strand breaks. The time evolution of the entire system reveals chemical pathways for carbonyl- and hydroxyl-hole formation in the sugar-moiety rings as well as base damage induced by $^\bullet$OH-radicals.

\noindent
The clustering method of ionization in the simulation of track-structure can be found in the literature, including in Refs. (Michalik 1993, Brenner 1992). The method is based on the search for energy-deposition in spherical-shaped clusters containing certain number of ionizations. For example the K-means clustering of ionization introduced in Ref. (Michalik 1993) focuses on the partitioning of set of N ionizations into K roughly spherical clusters (Michalik 1993). By fitting to empirical DSB yield, one may correlate the frequencies of occurrence of clusters with a given size and given number of ionizations to relative biological effects. Throughout such analysis it has been argued that locally multiply damaged sites are probably caused by energy depositions producing at least 2 to 5 ionizations localized, respectively, in sites of diameters of 1 to 4 nm (El Naqa 2012, Michalik 1993, Brenner 1992).

\noindent
A central assumption in these other approaches is the mathematical mapping between the number of ionizations localized in a cluster less than a critical size (sub-critical clusters) to a single site of damage in the track structure. In the track-structure simulation, if such a cluster is found, only one ionization inside the cluster is effectively considered to have caused the damage. All other ionizations inside that cluster are removed from further consideration, on the grounds that the site is already damaged. Mathematically this is equivalent of mapping all ionizations localized in a sub-critical cluster to a single damage, a homomorphism between the space of ionizations in the track-structure and the space of DNA damaged-sites. This is the phenomenon of saturation, discussed by Lea (Lea 1946), i.e., if multiple energy deposition sites occur very close together (e.g., at high LET) they become less biologically effective. In the approach presented in this work, we relaxed the above assumption because the occurrence of a single DNA lesion by ions in sub-critical clusters can be explicitly modeled in MD simulation.

\noindent
Furthermore, two beams of electrons and protons can be differentiated by the distribution of ions in the SVs and the type of DNA lesions they produce. Hence, we attempt to correlate the DNA lesions and the beam source by performing a first-principle calculation via hybridizing MC and MD simulations. Another effect not considered in previous works is the reaction of radical products with each other which could reduce the hydroxyl radical yield (Brenner 1992). In our MD simulation, we can easily trace the dynamics of $^\bullet$OH radicals. We demonstrated the interaction among $^\bullet$OH radicals within a cluster that leads to the formation of hydrogen-peroxides and network of OH$\cdots$OH hydrogen bonded molecules as well as recombination of $^\bullet$OH and H to form stable water that significantly lowers the reactivity of DNA-$^\bullet$OH (Abolfath 2011).

\noindent
Our study reveals a qualitative difference in the DNA damage induced by low energy electrons and protons. Electrons mainly generate small pockets of $^\bullet$OH-radicals, randomly dispersed in the SV volume. In contrast, protons generate larger clusters along a straight-line parallel to the initial direction of the particle. A quantitative comparison carried out on the DNA lesions in the limit of small track numbers shows that the total number of DSBs induced by a 1 MeV proton track is approximately four times greater than a 1 MeV electron track.
However, our preliminary observations indicate that this ratio may decrease because of the non-linearity on the dependence of the DSBs on the number of tracks, owing to the formation of the sub-critical clusters that occur with higher frequency as the number of tracks increases. From the track structures shown in Figs.\ref{Fig1} (a) and (b), one may expect that the onset of DSB saturation for protons is reached sooner than for electrons. Whether or not $\left\langle {\rm DSB}_{p} \right\rangle /\left\langle {\rm DSB}_{e} \right\rangle $approaches unity, the clinically reported RBE values (Paganetti 2003), requires further investigation.

\noindent
We note that in the current study we did not investigate the direct action, i.e., the direct ionization of DNA sugar-phosphate backbone, and the dissociation of water molecules into chemical species due to high energy events involving solvated electrons as discussed in (Karamitros 2011). We also considered only simulating hydroxyl free radicals and did not take into account the contribution of DNA damage induced by other free radical species.
Although the dissociation of water molecules into charged ions such as H$_3$O$^+$ and OH$^-$ are embedded in our simulation but as the current version of ReaxFF is not a suitable computational platform for processes involving solvated electrons, a full ab-initio method such as Car-Parinello MD (Car 1985, Hutter 2008) and QM-MM methods (Abolfath 2012) are also necessary.
We postpone our systematic study on these issues to forthcoming publications, however, spite of such limitation, we explored a number of interesting results using ReaxFF.

\noindent
In summary, we have developed a first-principle computational model to study the interaction of ionizing radiation with DNA molecules at the microscopic level.
As a general consensus, ionizing radiation changes the chemical environment of the DNA that subsequently leads to the chemical reactions in atomic level with manifestation in macroscopic scale, e.g., the DNA damage in the cellular level is the result of collective chemical reactions among chemical agents induced by ionizing radiation and DNA. As with any other “collective” phenomenon in physics, this problem can be studied in various scales, either at the macroscopic scale or in the atomic scale. In this study, we have taken the initial steps, as a proof-of-principle, to address the DNA-damage at the atomic level. Future works must be done to scale up the microscopic picture to the cellular level that experimental data are accessible.
The advantage of the hybrid MC modeling of ionizing radiation with Geant4-DNA and MD of DNA-molecules using ReaxFF is based on the feasibility in performing a multi-scale simulation of the cascade of events in physical and chemical pathways with biological endpoints of the irradiated cells, a complex system that comprises DNA molecules in reactive environment.

\section{Acknowledgement}
RMA thanks Nicole Ackerman, Satya Kumar, Mohammad Kohamdel, Siv Sivaloganathan and Adri van Duin for useful discussion.

\section{References}

\noindent 
{\bf Abolfath R M}, van Duin A C T, Brabec T 2011 Reactive Molecular Dynamics Study on the First Steps of DNA
Damage by Free Hydroxyl Radicals J. Phys. Chem. A \textbf{115}, 11045.
Animations and movies are available in:  http://qmsimulator.wordpress.com/

\noindent 
{\bf Abolfath R M}, Biswas P K, Rajnarayanam R, Brabec T, Kodym R, Papiez L 2012 Multiscale QM/MM Molecular Dynamics Study on the First Steps of Guanine-Damage by Free Hydroxyl Radicals in Solution, J. Phys. Chem. A \textbf{116}, 3940.

\noindent 
{\bf Aydogan B}, Bolch W E, Swarts S G, Turner J E, and Marshall D T 2008 Monte carlo simulations of site-specific radical attack to DNA bases, Radiat. Res. \textbf{169}, 223-231.

\noindent 
{\bf Antoniou D}, Abolfath R, and Schwartz S D, 2004 Transition Path Sampling Study of Classical Rate Promoting Vibrations, J. Chem. Phys. \textbf{121}, 6442.

\noindent %
{\bf Bagri A}, Mattevi C, Acik M, Chabal Y J, Chhowalla M, and Shenoy V B 2010, Nature Chem., {\bf 2}, 581.

\noindent 
{\bf Berendsen H J C}, van der Spoel D, van Drunen R 1996 Comp. Phys. Comm.  \textbf{91}, 43-56.

\noindent 
{\bf Brenner D J}, Ward J F 1992 Constraints on energy deposition and target size of multiply-damaged sites associated with DNA double strand breaks, Int. J. Radiat. Biol. \textbf{61}, 737-748.

\noindent 
{\bf Bernal M A} and Liendob J A 2009 An investigation on the capabilities of the PENELOPE MC code in nanodosimetry, Med. Phys. \textbf{36}, 620-625.

\noindent 
{\bf Bernal M A}, de Almeida C E, Sampaio C, Incerti S, Champion C, Nieminen P 2011 The invariance of the total direct DNA strand break yield'', Med. Phys. \textbf{38}, 4147- 4153.

\noindent 
{\bf Car R}, Parrinello M 1985 \prl, {\bf 55}, 2471.

\noindent 
{\bf Carlos S}, Oliveira  B,~and~ Oliveira-Brett A M 2012 In Situ DNA Oxidative Damage by Electrochemically Generated Hydroxyl Free Radicals on a Boron-Doped Diamond Electrode, Langmuir \textbf{28}, 4896-4901.

\noindent 
{\bf Chenoweth K}, van Duin A C T, and Goddard W A 2008, ReaxFF reactive force field for molecular dynamics simulations of hydrocarbon oxidation: Journal of Physical Chemistry A, {\bf 112}, 1040-1053.

\noindent 
{\bf El Naqa I}, Pater P and Seuntjens J 2012 Monte Carlo role in radiobiological modelling of radiotherapy outcomes, Phys. Med. Biol. \textbf{57}, 75.

\noindent 
{\bf Francis Z}, Incerti S, Capra R, Mascialino B, Montarou G, Stepan V, Villagrasa C 2011 Molecular scale track structure simulations in liquid water using the Geant4-DNA Monte-Carlo processes, Applied Radiation and Isotopes \textbf{69}, 220-226.

\noindent 
{\bf Friedland W}, Jacob P, Paretzke H G, Merzagora M and Ottolenghi A 1999 Simulation of DNA fragment distributions after irradiation with photons, Radiat. Environ. Biophys. \textbf{38}, 39.

\noindent %
{\bf Hutter J}, Ballone P, Bernasconi M, Focher P, Fois E, Goedecker S, Parrinello M, Tuckerman M E, 2008 CPMD code, version 3.13, MPI fuer Festkoerperforschung, Stuttgart IBM Zurich Research Laboratory.

\noindent %
{\bf Hess B}, van der Spoel D, Lindahl E 2008 J. Chem. Theory Comput., \textbf{4}, 435-447.

\noindent
{\bf Incerti S}, Baldacchino G, Bernal M, Capra R, Champion C, Francis Z, Gueye P, Mantero A, Mascialino B, Moretto P, Nieminen P, Villagrasa C, and Zacharatou C 2010a, The GEANT4-DNA project, Int. J. Model. Simul. Sci. Comput. {\bf 1}, 157–178.

\noindent
{\bf Incerti S}, Ivanchenko A, Karamitros M, Mantero A, Moretto P, Tran H N, Mascialino B, Champion C, Ivanchenko V N,  Bernal M A, Francis Z, Villagrasa C, Baldacchino G, Gueye P, Capra R, Nieminen P, and Zacharatou C 2010b, Comparison of GEANT4 very low energy cross section models with experimental data in water, Med. Phys. {\bf 37}, 4692–4708.

\noindent
{\bf Karamitros M}, Mantero A, Incerti S, Friedland W, Baldacchino G, Barberet P, Bernal M, Capra R, Champion C, EL Bitar Z, Francis Z, Gueye P, Ivanchenko A, Ivanchenko V, Kurashige H, Mascialino B, Moretto P, Nieminen P, Santin G,
Seznec H, Tran H N, Villagrasa C and Zacharatou C 2011, Modeling radiation chemistry in the Geant4 toolkit, Prog. Nucl. Sci. Tech. {\bf 2}, 503-508.

\noindent 
{\bf Lea D E} 1946 Actions of Radiations on Living Cells, (Cambridge University Press, Cambridge).

\noindent %
{\bf Lindahl E}, Hess B, van der Spoel D 2001 J. Mol. Model. 7, 306-317.

\noindent 
{\bf Kalantzis G}, Emfietzoglou D, Hadjidoukas P 2012 A unified spatio-temporal parallelization framework for accelerated Monte Carlo radiobiological modeling of electron tracks and subsequent radiation chemistry, Computer Physics Communications \textbf{183}, 1683.

\noindent %
{\bf Karplus M} and Petsko G A 1990 Molecular dynamics simulations in biology, Nature \textbf{347}, 631-639.

\noindent %
{\bf Karplus M} and McCammon J A 2002 Molecular dynamics simulations of biomolecules, Nature Structural Biology \textbf{9}, 646-652.

\noindent 
{\bf McNamara A L}, Guatelli S, Prokopovich D A, Reinhard M I and Rosenfeld A. B. 2012 A comparison of X-ray and proton beam low energy secondary electron track structures using the low energy models of Geant4, International Journal of Radiation Biology, \textbf{88}, 164-170.

\noindent 
{\bf Michalik V} 1993 Rad.  Res. \textbf{134}, 265-270;
Verhaegena E, and Renier B 2004 Rad.  Res. \textbf{162}, 592-599.

\noindent 
{\bf Moiseenko V V}, Hamm R N, Waker A J, and Prestwich W V 1998 Modelling DNA damage induced by different energy photons and tritium beta-particles, Int. J. Radiat. Biol. \textbf{74}, 533-550.

\noindent
Monti S, van Duin A C T, Kim S-Y, Barone V 2012, Exploration of the Conformational and Reactive Dynamics of Glycine and Diglycine on TiO2: Computational Investigations in the Gas Phase and in Solution, J. Phys. Chem. C {\bf 116}, 5141-5150.

\noindent 
{\bf Mozumder A} 1985 Early Production of Radicals from Charged Particle Tracks in Water, Radiat. Res. \textbf{104}, 33-39.

\noindent 
{\bf Munteanu M G}, Vlahovicek K, Parthasaraty S, Simon I and Pongor S 1998 Rod models of DNA: sequence-dependent anisotropic elastic modeling of local bending phenomena, Trends Biochem. Sci. {\bf 23}, 341-346;
http://hydra.icgeb.trieste.it/dna/model\_it.html

\noindent 
{\bf Nikjoo H}, and Charlton D E, 1995 Calculation of range and distributions of damage to DNA by high- and low-LET radiations in Radiation Damage to DNA: Structure/Function Relationships at Early Times (A. F. Fuciarelli and J. D. Zimbrick, Eds. Battelle Press, Columbus).

\noindent %
{\bf Nikjoo H}, O'Neill P, Terrissol M, and Goodhead D T 1994 Modeling of radiation-induced DNA damage: the early physical and chemical event, Int. J. Radiat. Biol. \textbf{66}, 453-457.

\noindent %
{\bf Nikjoo H}, O'Neill P, Goodhead D T and Terrissol M 1997 Computational modelling of low-energy electron-induced DNA damage by early physical and chemical events, Int. J. of Radiat. Biol. \textbf{71}, 467-483.

\noindent %
{\bf Paganetti H}, Niemierko A, Ancukiewicz M, Gerweck L E, Goitein M, Loeffler J S, Suit H D 2003
Int J Radiat. Oncol. Biol. Phys., {\bf 53}, 407.

\noindent 
{\bf Plimpton S J} 1995 J. Comp. Phys. {\bf 117}, 1.

\noindent %
{\bf Plimpton S J}, Pollock R, Stevens M; Ewald P -M, 1997 in Proc of the Eighth SIAM Conference on Parallel Processing for Scientific Computing, Minneapolis, MN.

\noindent 
{\bf Roots R} and Okada S 1975 Estimation of life times and diffusion distances of radicals involved in x-ray-induced DNA strand breaks of killing of mammalian cells, Radiat Res. \textbf{64}, 306-320.

\noindent %
{\bf Schwartz S D} and Schramm V L 2009 Enzymatic transition states and dynamic motion in barrier crossing, Nature Chemical Biology \textbf{5}, 551-558.

\noindent 
{\bf Semenenko V A}, Stewart R D, Ackerman E J, 2005 Monte Carlo Simulation of Base and Nucleotide Excision Repair of Clustered DNA Damage Sites. I. Model Properties and Predicted Trends, Radiat. Res. \textbf{164}, 180-193; \textit{ibid.} 2005 Comparisons of Model Predictions to Measured Data.\textbf{164}, 194-201.

\noindent 
{\bf Stewart R D}, Yu V K, Georgakilas A G, Koumenis C, Park J H and Carlson D J 2011 Effects of Radiation Quality and Oxygen on Clustered DNA Lesions and Cell Death, Radiat. Res., \textbf{176}, 587-602.

\noindent 
{\bf Schneider G} and Fechner U, 2005 Computer-based de novo design of drug-like molecules, Nature Reviews Drug Discovery \textbf{4}, 649-663; Durrant J D and McCammon J A, 2011 Molecular dynamics simulations and drug discovery BMC Biology \textbf{9}, 71.

\noindent 
{\bf Sies H} 1993 Strategies of antioxidant defense, Europ. J. Biochem. \textbf{215}, 213.

\noindent 
{\bf Terrisol M}, and Beaudre A 1990 Simulation of space and time evolution of radiolytic species induced by electrons in water, Radiat. Prot. Dosimetry \textbf{31}, 171.

\noindent %
{\bf Van Der Spoel D}, Lindahl E, Hess B, Groenhof G, Mark A E, Berendsen H J C 2005 J. Comput. Chem. \textbf{26}, 1701-1718.

\noindent 
{\bf van Duin A C T}, Dasgupta S, Lorant F, Goddard W A 2001 J. Phys. Chem. A \textbf{105}, 9396-9409.

\noindent 
{\bf Wilson W E}, and Paretzke H G 1994 A stochastic model of ion track structure, Radiat. Prot. Dosimetry \textbf{52}, 249.

\noindent 
{\bf Ward J F} 1988 DNA damage produced by ionizing radiation in mammalian cells: identities, mechanisms of formation, and reparability, Prog. Nucleic Acid Res. Mol. Biol. \textbf{35} 95-125.

\noindent 
{\bf Zaider M}, and Brenner D J 1984 On the stochastic treatment of fast chemical reactions, Radiat. Res., \textbf{100}, 245-256.



\end{document}